\documentclass[twocolumn]{aa}

\usepackage{amsmath}
\usepackage{txfonts}
\usepackage{natbib}
\usepackage{graphics}
\usepackage{graphicx}
\usepackage{epsfig}
\usepackage{amssymb}
\usepackage{xspace}

\usepackage{epsf,psfig}
\usepackage[dvips]{color}
\usepackage[latin1]{inputenc}
\usepackage[T1]{fontenc}
\usepackage{color}
\definecolor{red}{rgb}{0.7,0,0}
\definecolor{blue}{rgb}{0,0,0.7}

\newcommand{\un}[1]{~\hspace{-1pt}\ensuremath{\mathrm{#1}}}

\newcommand{\integ}{{\it INTEGRAL}\xspace}
\newcommand{\ibis}{IBIS\xspace}

\newcommand{\isgri}{ISGRI\xspace}
\newcommand{\picsit}{PICsIT\xspace}
\newcommand{\jemx}{JEM-X\xspace}

\newcommand{\coma}{Coma~Cluster\xspace}

\def\m{$^\prime$}
\def\eg{{\it e.g. }}
\def\etal{et~al.~}
\def\ie{{\em i.e. }}
\newcommand{\gammaray}{$\gamma$-ray\xspace}

\newcommand{\xray}{X-ray\xspace}

\begin{document}

\title{Imaging extended sources with coded mask telescopes: Application to the INTEGRAL IBIS/ISGRI instrument}
\author{M. Renaud\inst{1,2}, A. Gros\inst{1}, F. Lebrun\inst{1,2}, R. Terrier\inst{2},
A. Goldwurm\inst{1,2}, S. Reynolds\inst{3}, E. Kalemci\inst{4,5}}

\offprints{M. Renaud : mrenaud@cea.fr}

\institute{Service d'Astrophysique, CEA-Saclay, 91191,
Gif-Sur-Yvette, France \and APC-UMR 7164, 11 place M. Berthelot,
75231 Paris, France \and Department of Physics, NC State
University, 2700 Stinson Drive, Box 8202, Raleigh, NC 27695, USA
\and Space Sciences Laboratory, 7 Gauss Way, University of
California, Berkeley, CA 94720-7450, USA \and Sabanc\i~University,
Orhanl\i~- Tuzla, Istanbul, 34956, Turkey}

\date{Received 8 March 2006 / Accepted 30 May 2006}
\authorrunning{M. Renaud et al.}
\titlerunning{Extended sources through the \ibis coded mask telescope}

\abstract{In coded mask techniques, reconstructed sky images are
pseudo-images: they are maps of the correlation between the image recorded
on a detector and an array derived from the coded mask pattern.}
{The \integ/\ibis telescope provides images where the
flux of each detected source is given by the height of the local peak
in the correlation map. As such, it cannot provide an estimate of
the flux of an extended source. What is needed is intensity sky images
giving the flux per solide angle as typically done at other wavelengths.}
{In this paper, we present the response of the \integ \ibis/\isgri coded
mask instrument to extended sources. We develop a general method
based on analytical calculations in order to measure the intensity and
the associated error of any celestial source and validated with Monte-Carlo
simulations.}{We find that the sensitivity degrades almost linearly with
the source extent. Analytical formulae are given as well as an easy-to-use
recipe for the \integ user. We check this method on \ibis/\isgri data but
these results are general and applicable to any coded mask telescope.}{}


 \maketitle


\section{Introduction}
\label{s:introduction}

The usual techniques of imaging the sky by focusing the radiation become
ineffective above $\sim$ 15\un{keV} due to technological constraints. Future
observatories (\eg Constellation-X/HXT, Hailey \etal 2004 ; Simbol-X, Ferrando \etal 2005)
will push this limit up to 60-70\un{keV} with the use of multi-layer grazing optics \cite{c:ramsey02}.
Collimators using the standard on/off techniques are limited by the source density
at low energies (< 50\un{keV}) and by the internal background variability at
higher energies. Both problems are addressed with coded-aperture techniques, which
have been employed successfully in high-energy astronomy in several previous
missions such as the GRANAT/SIGMA instrument \cite{c:paul91} or the BeppoSAX
Wide Field Cameras \cite{c:jager97}. In such telescopes, source
radiation is spatially modulated by a mask consisting of an array of
opaque and transparent elements and recorded by a position sensitive
detector, allowing simultaneous measurement of source plus background
in the detector area corresponding to the mask holes and background
only in that corresponding to the opaque elements. For each
point source in the field of view the two-dimensional distribution of
events projected onto the detector reproduces a unique shadow of the mask
pattern and the recorded shadowgram is the sum of many such
distributions. The popular way to reconstruct the sky image
is based on a correlation procedure between the recorded image and a
decoding array derived from the spatial characteristics of the mask
pattern. One can describe the usual properties of telescopes for
these systems as follows: the angular resolution, defined by the angle
subtended by the smallest mask element seen from the detector, and the
field of view (FOV) divided in two parts: the Fully Coded Field of
View (FCFOV) representing the sky region where the recorded
source radiation is fully modulated by the mask and the Partially
Coded Field of View (PCFOV) the region where a source projects only a
part of the mask pattern on the detector.

Many mask patterns have been designed and employed in the
past for optimizing the imaging quality of point-like sources in the
high-energy domain. Recently, Sch\"afer (2004) has presented a novel
method for imaging extended sources by constructing mask patterns with
a pre-defined Point Spread Function (PSF). In this paper, we have
developed a method to reconstruct the intensity and estimate the
corresponding error of extended sources seen through a given coded
mask telescope. This work is motivated by the fact that several
sources of interest in \gammaray astronomy must be considered as
extended in order to properly derive their intensity such as supernova
remnants (hereafter, SNRs): the Tycho SNR (with an apparent diameter $\theta$ of
$\sim$ 8\m), SN~1006 ($\theta \sim$ 30\m), G~347.3-0.5 ($\theta \sim$ 1$^{\circ}$) or
RX~J0852-4622 ($\theta \sim$ 2$^{\circ}$), clusters of galaxies (\eg
\coma, $\theta \sim 1^{\circ}$), and diffuse interstellar emission from
various high-energy processes. We have applied this method to the
\ibis/\isgri instrument onboard the \integ satellite but the results
presented here could be easily adapted to any coded mask telescope.
We briefly describe the basic properties of \ibis and the principles
of the standard data analysis we used for simulating celestial extended
sources in section \ref{s:method}, considering the simplest case of a
uniform disk. We present our results concerning the \ibis response for
this source geometry over a large range of radii (section \ref{s:disk})
as well as in an astrophysical case: the supernova remnant SN~1006 (section
\ref{s:SN1006}). These results allow us to find a general method to
reconstruct the flux and the associated error of an extended source
presented in sections \ref{s:real_flux} and \ref{s:real_err}, respectively.
We also describe the tests we performed on the Crab Nebula and compare
the results of this method to those of the standard analysis (section
\ref{s:crab}). In a joint paper, we have successfully applied this method
on the \ibis/\isgri data of the \coma \cite{c:renaud06}, the first extended
source detected with \ibis. Finally, we give practical tips on the use of
this method for any interested observer.


\section{Imaging sources with a coded mask telescope}
\label{s:method}

\subsection{General imaging procedure}
\label{s:general}

In coded aperture imaging systems, a mask consists in an array M of 1
(transparent) and 0 (opaque) elements. The detector array D (the real
image) is simply the convolution of the sky image S with M plus an
unmodulated background array B. If there exists an inverse correlation
array G such that (M$\star$G)$_{ij} = \delta_{ij}$, then the reconstructed sky
image is given by the following operation \cite{c:fenimore81}:

\begin{equation}
S' = D \star G = S \star M \star G + B \star G = S + B \star G
\end{equation}

In the case where the mask M is derived from a cyclic replication of
the same basic pattern and the background B is given by a flat array
then the term B$\star$G is constant and can be removed. The quality of
the object reconstruction therefore depends on the choice of M and G
\cite{c:caroli87}. Fenimore \& Cannon (1978) have found mask patterns
with such properties, including the Modified Uniformly Redundant
Arrays (hereafter, MURAs). These coded masks are nearly-optimum and
possess a correlation inverse matrix by setting G = 2M - 1. For those
systems the resulting sky image of a single point-like source in the
FCFOV will have a main peak at the source position of roughly
the size of one projected mask element, with flat sidelobes in the
FCFOV (for a perfect detector) and 8 main source ghosts in the PCFOV.

\subsection{\ibis properties}
\label{s:ibis}

We consider the \ibis telescope \cite{c:ubertini03} launched
onboard the ESA \gammaray space mission \integ \cite{c:winkler03} on
October 2002. The \ibis coded mask imaging system includes a replicated
MURA (see section \ref{s:general}) mask \cite{c:gottesman89} of tungsten
elements and 2 pixellated \gammaray detectors: \isgri \cite{c:lebrun03},
the low-energy camera (15\un{keV}~-~1\un{MeV}) and \picsit, operating
between 175\un{keV} and 10\un{MeV} \cite{c:labanti03}. We focus here on
\ibis/\isgri but the results can be adapted to any coded mask imaging system.
The physical characteristics of the \ibis telescope define a FCFOV of
8.3$^{\circ} \times$ 8.3$^{\circ}$ and a total FOV of 29$^{\circ}$
$\times$ 29$^{\circ}$ at zero sensitivity. The nominal angular resolution
is 12\m~(FWHM) and standard \isgri images are sampled every $\sim 5$\m~, which
is the angular size of detector pixels seen from the mask, allowing fine imaging and
positioning of detected celestial sources. The decoding array G is oversampled
at the pixel scale in order to optimize the signal to noise ratio of point-like
sources in the reconstructed image. One may refer to the papers of Gros \etal
(2003) and Goldwurm \etal (2003) for details on the shape of the System Point
Spread Function (hereafter, SPSF) and the full description of the implemented
algorithm for \ibis data, respectively. Goldwurm et al. (2003) show a variation
of finely balanced cross correlation \cite{c:fenimore81} generalized to the
total (FC+PC)FOV.

\subsection{Imaging extended sources}
\label{s:imaging}

In the following, any extended source will be considered as a large
number of unresolved points. The principle of our method for imaging
extended sources is as follows: first, we have calculated the fraction
of pixel area covered by a given projected mask element (called Pixel
Illumation Factor, PIF) corresponding to each point-like source given
its relative sky position. These calculations take into account all the
instrumental features of \ibis such as efficiency, dead zones between pixels
and detector modules of \isgri, mask thickness and transparencies of all the
intervening materials in the ISGRI energy range (15\un{keV} -- 1\un{MeV}).
Second, we have performed a weighted sum of each of the point-like source
contributions to obtain the shadowgram of any extended sky source according
to the following equation:

\begin{equation}\label{e:convol}
D_{kl} = \sum_{w=1}^{N}f_{w}\times{PIF_{kl}^{w}}
\end{equation}

where D$_{kl}$ is the image recorded on \isgri, N refers to the
number of unresolved points forming the extended source, and f$_{w}$
represents the flux of each point source (w) with $\sum{f_{w}} = F$, the
total flux of the extended source.

\begin{figure}[htb]
\centering
\includegraphics[width=0.97\columnwidth]{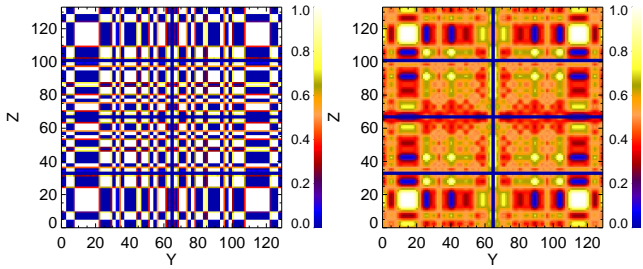}

\caption{The recorded \isgri shadowgram of the PIF values
for an on-axis point-like source (left) and that of a uniform disk
of $\sim 20$\m~radius (right) without background. Dead zones
between modules and mask elements are clearly visible.}

\label{f:Shadowg_OnAxis}
\end{figure}

In order to illustrate the blurring effect due to the source extent,
Figure \ref{f:Shadowg_OnAxis} shows the \isgri shadowgram of PIF values
corresponding to an on-axis point-like source and that of a uniform disk of $\sim$
20\m~radius. The overall pattern of the PIF shadowgram has a very little
dependence on energy and only the relative amplitude of fluctuations depends on
the energy. Finally, in order to reconstruct the sky images, we have used the standard
deconvolution included in the Off-Line Scientific Analysis \cite{c:goldwurm03},
version 5.0. We decided to perform these simulations in connection with what
an observer can expect by analyzing \ibis/\isgri data of any celestial region.
We have chosen to simulate a disk as discussed in section \ref{s:introduction},
assuming that each individual point source contribution forming
the extended source emit the same flux (\ie the factor f$_{w}$ in the equation
\ref{e:convol} is constant, each extended source is uniform). We have
distinguished two different relative positions in the sky:
an on-axis source and one in the PCFOV, 9$^{\circ}$ away from the telescope
axis, \ie where the \ibis sensitivity is reduced by a factor of almost 2.


\section{Sensitivity to extended sources}
\label{s:sensi_exten}

Simulations were performed for several radii, from 0 (\ie, a
point-like source) to $\sim$ 1$^{\circ}$ \textit{keeping constant the
total flux} F equal to $\sum_{w=1}^{N}f_{w}$. One should
notice that in the standard reconstructed images, the flux of
point-like sources in the FOV is given by the peak of their associated
SPSF after fitting with a two-dimensional gaussian (see Gros \etal
2003). Therefore, for each reconstructed sky image, we have measured
the maximal pixel flux within the apparent diameter of each extended
source (hereafter, f$_{max}$) to estimate the \ibis/\isgri response to
the global peak reconstruction.

\subsection{Uniform disk}
\label{s:disk}

\begin{figure}[htb]
\centering
\includegraphics[width=0.85\columnwidth]{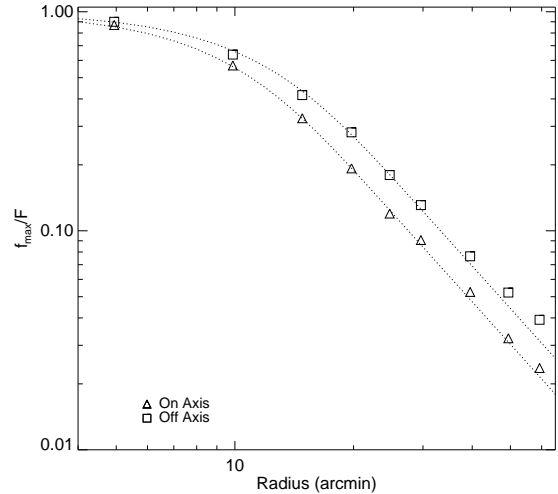}

\caption{The relative reconstructed peak flux as a function of the
radius of the uniform disk for two different positions: on-axis
(triangles) and 9$^{\circ}$ away from the pointing (labeled
off-axis, squares). The two curves have the same behavior and
present two main features: below $\sim 8'$ radius, the loss in
flux is limited whereas above this value, the flux decreases as
R$^{-2}$. The dotted lines give the expected behavior based on the
equation \ref{e:disk}.}

\label{f:Flux_DISK}
\end{figure}

This case could relate to the Crab-like SNRs (plerions or pulsar-wind
nebulae) containing pulsars that inject a relativistic wind of
synchrotron-emitting plasma into the surrounding medium. These pulsar-wind
nebulae can be seen in first approximation as uniform disks. Figure
\ref{f:Flux_DISK} presents f$_{max}$ as a function of the radius for the
two different source positions in the \ibis FOV.

The two curves present different behaviors: below a radius of $\sim$ 8',
the extent of the source is still comparable to the \ibis angular
resolution and the flux loss is less than 30 \%. For more extended
sources, the SPSF differs from that of a point-like source and the
flux decreases as R$^{-2}$ (and R$^{-1}$ in the case of a uniform ring) due
to the simple flux dilution. One can easily show that this is well described
by the equation:

\begin{equation}
\label{e:disk}
\frac{f_{max}}{F} = 2\frac{\sigma_{spsf}^{2}}{R^{2}}\times{(1 - e^{-R^{2}/2\sigma_{spsf}^{2}})}
\end{equation}

where $\sigma_{spsf}$ is the width of the SPSF and R the source radius. The
difference between the on-axis and off-axis curves is coming from the
fact that the width of the \ibis SPSF is not constant over the relative
positions in the FOV: we have fitted the SPSF for the two positions with
a bi-dimensional gaussian and found that the width in the on-axis case is
close to the nominal value, $\sim$ 14' (FWHM) and is around 17.5' in the
off-axis one. These results are similar to those of Gros \etal (2003,
Fig. 2) and explain what we observe in Figure \ref{f:Flux_DISK}: the
reconstructed peak flux of an extended source depends less sensitively to
the source size when the width of the SPSF is larger. We will discuss in
details the characteristics of the \ibis SPSF in section \ref{s:real_flux}.

\subsection{An astrophysical case: SN~1006}
\label{s:SN1006}

\begin{figure}[htb]
\centering
\includegraphics[width=0.93\columnwidth]{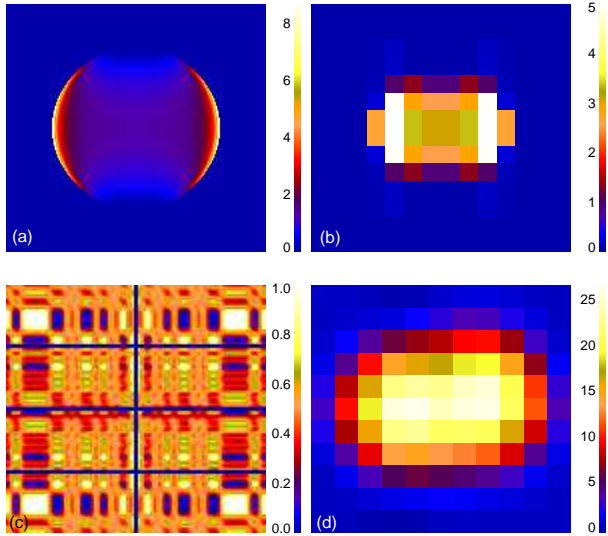}

\caption{(a) The expected image at 28\un{keV} (in units of 10$^{-26}$
ergs cm$^{-2}$ s$^{-1}$ Hz$^{-1}$ sr$^{-1}$) based on simulations performed
by Reynolds (1999). (b) The sampled sky image in the \isgri sky pixels size
(in units of ct s$^{-1}$) . (c) The corresponding shadowgram of the PIF values
and (d) the reconstructed image after the standard deconvolution 
(in units of ct s$^{-1}$).}

\label{f:SN1006_Ima}
\end{figure}

SN~1006 is an historical Galactic SNR of $\sim 30'$ diameter, observed
with \integ for $\sim$ 1 Msec in the first two years of the mission.
Reynolds (1999) has modelled the predicted hard \xray emission of this
SNR with two components: the synchrotron radiation, concentrated in two
bright limbs and based on the calculations of Reynolds (1998), and a
non-thermal bremsstrahlung component more symmetrically distributed.
At energies below about 100\un{keV}, synchrotron emission will dominate.
Figure \ref{f:SN1006_Ima} presents four images of this SNR, with the expected
morphology of SN~1006 at 28\un{keV}, the sampled image in the \isgri sky
pixels, and the results of our simulations.  In this example, SN~1006
is located in the FCFOV.

The results of the analysis of the \jemx and \ibis/\isgri data of SN~1006
are presented in Kalemci \etal (2006). The source was not detected with
\isgri.  Using the same method as in section \ref{s:disk},
it was possible to extract an upper limit on the hard \xray emission of
SN~1006. Since the expected morphology is almost symmetric and the emission
is concentrated in two extended limbs, the ratio between the maximal value
in the reconstructed image at the position of each limb and their respective
global flux provides a correction factor. In Kalemci \etal (2006), the
"standard" upper limits (obtained as if the SN~1006 is an unresolved source)
were divided by this correction factor of $\sim$ 0.7 to take the flux dilution
into account. Note that a relative reconstructed peak flux of 0.7 is close
to what we obtained for an extended source of $\sim 10'$ radius (see Fig.
\ref{f:Flux_DISK}).


\section{General method}
\label{s:gen_method}

We have shown that any coded mask imager such as \ibis works with
extended sources just as predicted based on its SPSF properties:
the reconstructed peak flux follows the simple flux dilution over the
large number of unresolved points constituting an extended
source. Moreover, as previously presented by Gros \etal (2003), the
width of the \ibis SPSF varies within the FOV.

\subsection{Reconstructed flux of extended sources}
\label{s:real_flux}

As described in section \ref{s:sensi_exten}, the flux of a point-like
source in a reconstructed \ibis image is given by the peak of the SPSF.
Therefore, as shown in the previous sections, this will give a biased
estimation of the flux of an extended source. For proper flux
estimation, one needs to convert the reconstructed standard images into
images of intensity (\ie flux per solid angle or sky pixel), as is typical
in astronomy. In that case, the global flux of any source (point-like or
extended) would be given by integrating intensities over the source extent.
The method is to divide standard flux images by the SPSF effective area (the
solide angle subtended by the SPSF on the celestial sphere, \ie the ratio
between its integral and the peak value) but the main difficulty is that the
shape of the \ibis SPSF varies within the FOV, so it is necessary to construct
a map of its effective area (hereafter, $\tilde{A}$).

\begin{figure*}[htb]
\centering
\includegraphics[scale=0.75, angle=0]{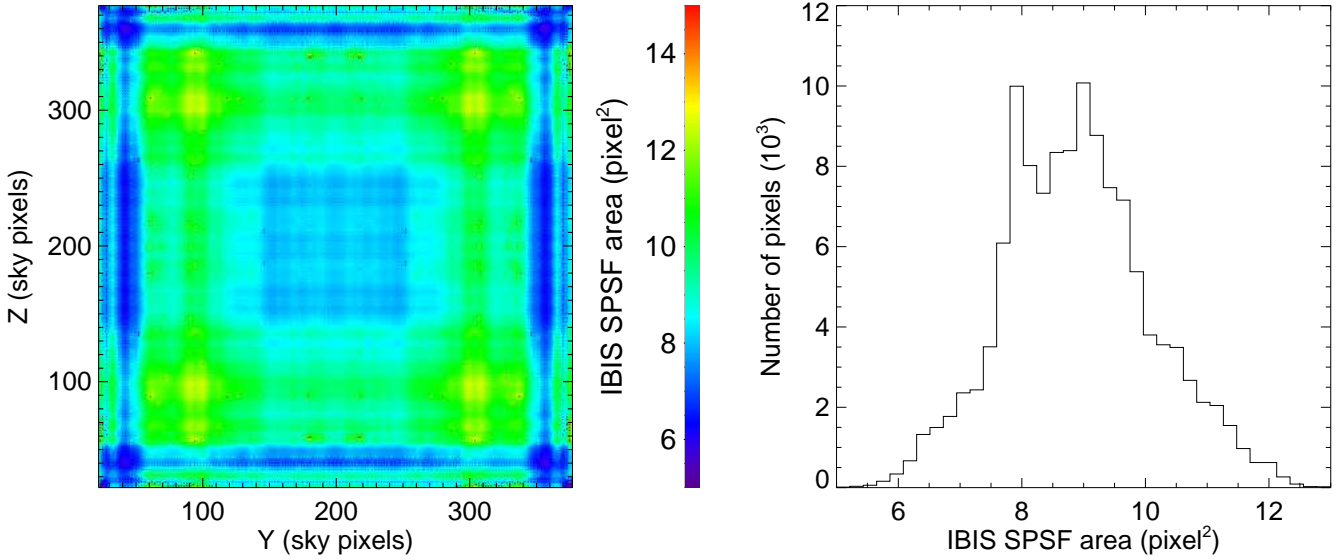}

\caption{Map of the \ibis SPSF effective area $\tilde{A}$ in a
linear scale (left) and the histogram of its values (right).}

\label{f:Width_Sum}
\end{figure*}

We have calculated point-like sources each $\sim 15'$ (3 \isgri sky
pixels sampling) in the first quarter of the \ibis FOV with a fixed flux
and fitted each corresponding reconstructed image at the source position
by a 2D gaussian. We assume that these images must have a central
symmetry due to that of the MURA coded mask and then we have projected
each quarter of images where we performed the simulations on the three
others according to the central symmetry. Figure \ref{f:Width_Sum}
presents the map of $\tilde{A}$ of the \ibis SPSF and the histogram of
its values. The map is almost flat in the FCFOV and the values globally
increase in the PCFOV. Its histogram shows an appreciable dispersion
($\sigma$ $\sim$ 1.3 sky pixel$^{2}$) with a mean value of 8.9 sky pixel$^{2}$.
To obtain reconstructed sky images in intensity, one can then apply the
following transformation:

\begin{equation}
\label{e:transfo_flux1}
I'_{mn} = \frac{S'_{OSA}}{\tilde{A}_{mn}}
\end{equation}
\begin{equation}
\label{e:transfo_flux2}
I'_{S}  = \sum_{\Omega}I'_{mn}
\end{equation}

where S'$_{OSA}$ is the reconstructed flux image obtained with the
standard deconvolution. The legitimacy of such a transformation is shown in
Appendix \ref{a:flux} and is based on the sole assumption that $\tilde{A}$
does not change on scales smaller than the SPSF width. Thus, for extended sources
simulated as described in section \ref{s:method}, we have built these intensity
images $I'_{mn}$. For each of them we have integrated over the source extent as
seen by IBIS ($\Omega$, \ie the physical size of the source convolved with the
\ibis SPSF $\sim$ source radius plus 3 $\times \sigma_{spsf}$) and we have checked
that this final value corresponds to the input global flux of our simulations
with a dispersion of less than 10 \%.

\subsection{Tests on the Crab Nebula observations}
\label{s:crab}

\begin{figure}[htb]
\centering
\includegraphics[width=0.495\columnwidth]{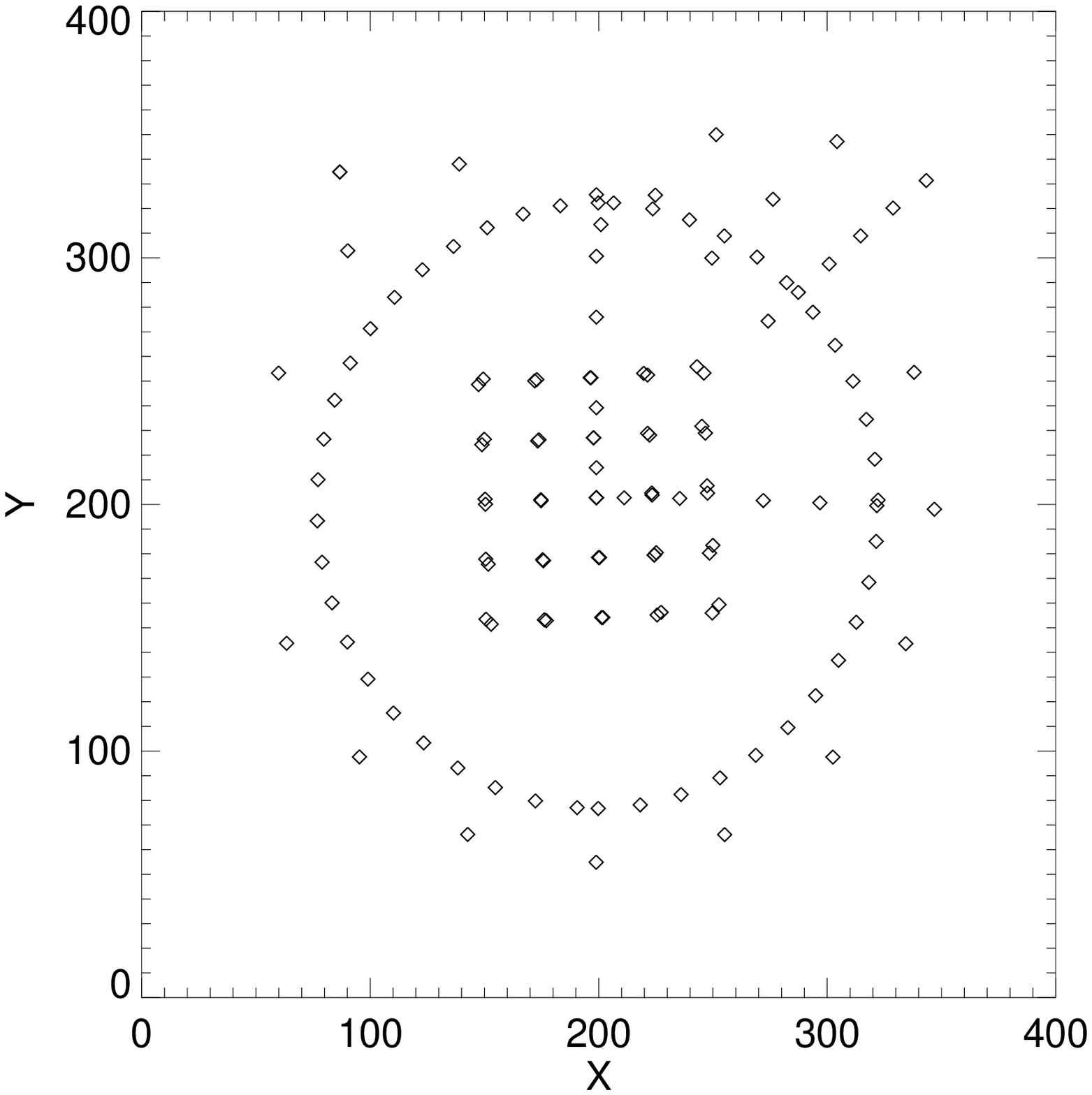}
\includegraphics[width=0.495\columnwidth]{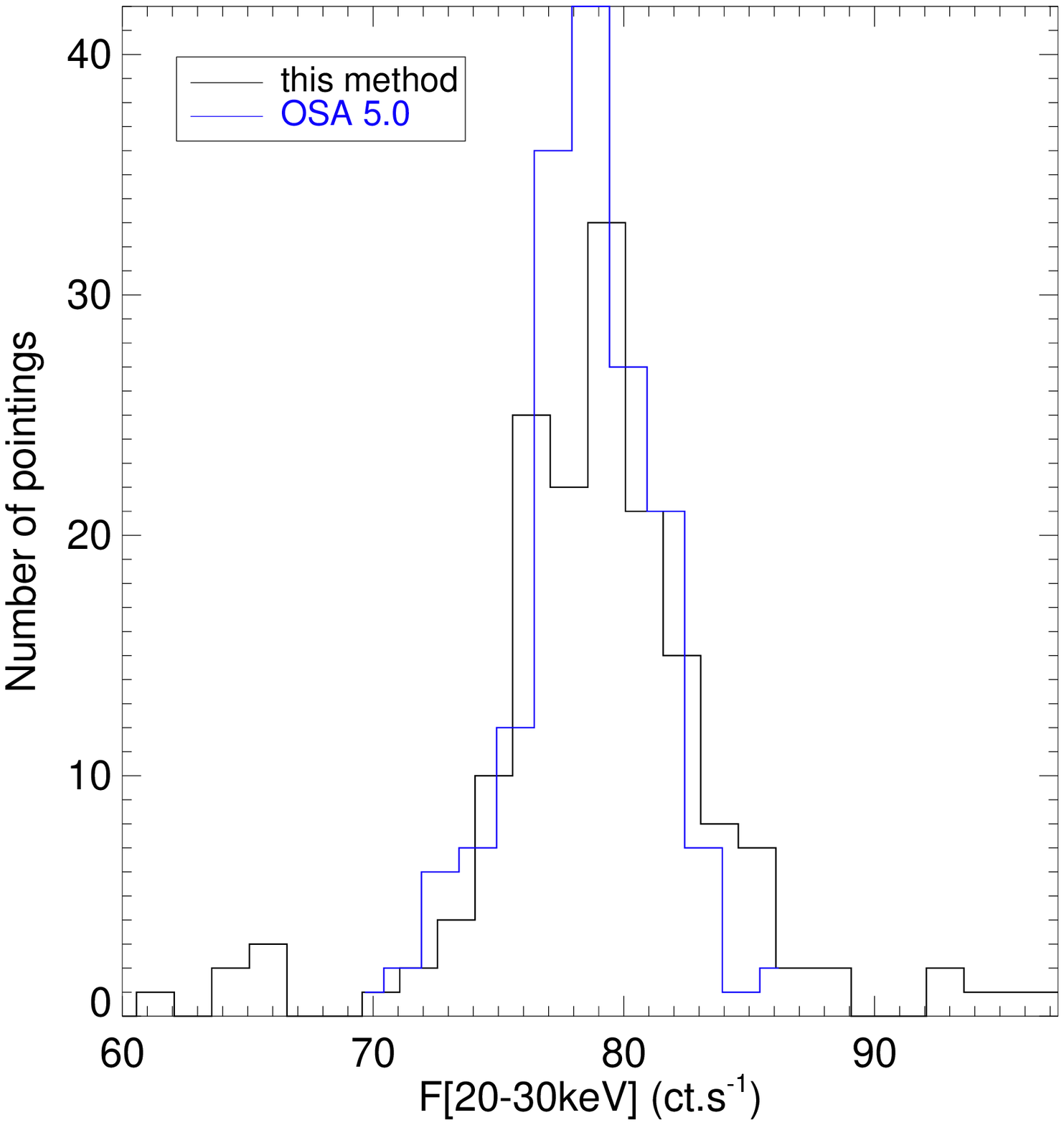}

\caption{Left: Relative positions of the Crab Nebula for the 164
analyzed pointings during the rev. 102, 170. Right: Histograms of
the reconstructed fluxes of the Crab with the standard analysis
OSA 5.0 (in blue) and with our method (in black)}

\label{f:crab_obs}
\end{figure}

We have also tested this method on Crab Nebula observations. This source
is considered as point-like for \ibis. We used data obtained during the
revolutions 102, 170 and 300 containing 164 science windows (individual
pointings, hereafter scws) and applied the same procedure as described above.
Figure \ref{f:crab_obs} presents the relative positions of the Crab for the
analyzed scws and the histograms of the fluxes found with our method and the
standard one in the 20-30\un{keV} energy range. Even if our method shows a larger
dispersion in the reconstructed Crab fluxes ($\sigma_{method}$ $\sim$ 3.1 s$^{-1}$
and $\sigma_{OSA 5.0}$ $\sim$ 2.3 s$^{-1}$), the mean values are
compatible (<F$_{method}$> $\sim$ 78.9 s$^{-1}$ and <F$_{OSA 5.0}$>
$\sim$ 78.7 s$^{-1}$). Since the Crab Nebula is an unresolved source for
\ibis, the reconstructed flux with our method depends on the source location
since $\tilde{A}$ varies with the position in the field of view. However, the
sensitivity of this dependence decreases with the source extent.


\subsection{Estimation of the associated errors}
\label{s:real_err}

Although the flux of any extended source can be measured with the
method described above, the estimation of the corresponding errors
is not simple. In fact, the reconstructed counts in the sky pixels
within the SPSF of the \ibis telescope are correlated. Since we
propose to sum the fluxes previously divided by $\tilde{A}$ over
the size of the source, the associated variance should be 
$\sum{\sigma'^{2}_{mn}/\tilde{A}^{2}_{mn}}$ ($\sigma'^{2}_{mn}$ is
the OSA flux variance) plus a covariance term. The latter depends
on the position in the FOV and the source extent.

We have developed the standard equations of the convolution in Appendix
\ref{a:variance} to find a simple analytical expression of the variance
associated to the equations \ref{e:transfo_flux1} and \ref{e:transfo_flux2} such that:

\begin{equation}
\label{e:transfo_var1}
\sigma'^{2}_{I'_{S}} \approx \sigma^{2}_{B,ct.s^{-1}} \sum_{\Omega}\frac{1}{\tilde{E}_{mn}\tilde{A}_{mn}}
\end{equation}
\begin{equation}
\label{e:transfo_var2}
\sigma'^{2}_{I'_{S}} \approx \sum_{\Omega} \frac{\sigma'^{2}_{OSA}}{\tilde{A}_{mn}} \approx N_{spsf} <\sigma'^{2}_{OSA}>
\end{equation}

\noindent where $\sigma^{2}_{B,ct.s^{-1}}$ is the detector variance of the background
count rate (in units of count s$^{-2}$), ${E}_{mn}$ is the imaging efficiency and
$<\sigma'^{2}_{OSA}>$ is the OSA flux variance averaged over the region of pixel
fluxes summation $\Omega$. N$_{spsf}$ refers to the number of SPSFs within $\Omega$.
Equation \ref{e:transfo_var2} can be interpreted as follows: the width of the SPSF
($\sigma_{spsf}$) is the correlation length of the instrument beyond which sky pixels are
uncorrelated. Therefore, at these scales, the final variance associated with the
reconstructed flux of an extended source by summing the pixel fluxes is simply given
by the sum of the variances. This can be then approximated to the mean variance times
the number of SPSFs within $\Omega$ (see Appendix \ref{a:practical_tips}).

In order to check this analytical estimation of the associated error, we
have performed Monte-Carlo simulations of extended sources seen through the
\ibis coded mask as a sum of unresolved points. Shadowgrams are given by
equation \ref{e:convol} for which we have added a constant background
term. The total flux of each extended source was fixed at 50 s$^{-1}$ and
that of the background at 10$^{3}$ s$^{-1}$. 500 shadowgrams are simulated for
different source sizes (assumed to be uniform disks), from 0$^{\circ}$ (point-like)
to 1$^{\circ}$ radius, following the equation: $\sqrt{D_{kl}}\times{N(0,1)} + D_{kl}$
where D defines the \isgri expected count rate shadowgram at the pixel (k,l). N(0,1)
is the normal distribution. Each of these simulated shadowgrams was deconvolved using the
standard method of the \ibis data analysis in order to obtain 500 reconstructed images in flux,
variance and significance. We performed these calculations for three different
source positions in the FOV: one on-axis (Y=200.5, Z=200.5) and two in the PCFOV
at $\sim$ 9.5$^{\circ}$ (Y=300.5, Z=250.5, labelled ``Off Axis1'') and
$\sim$ 13.5$^{\circ}$ near the edge of the FOV (Y=120.5, Z=50.5, labelled
``Off Axis2''). These calculations allow us to control the statistics for any
source extent and position in the FOV. We have reconstructed the source fluxes
using the method described in section \ref{s:real_flux} and measured the width of
each flux histogram, which corresponds to the error associated to this method.

\begin{figure}[htb]
\centering
\includegraphics[width=0.9\columnwidth]{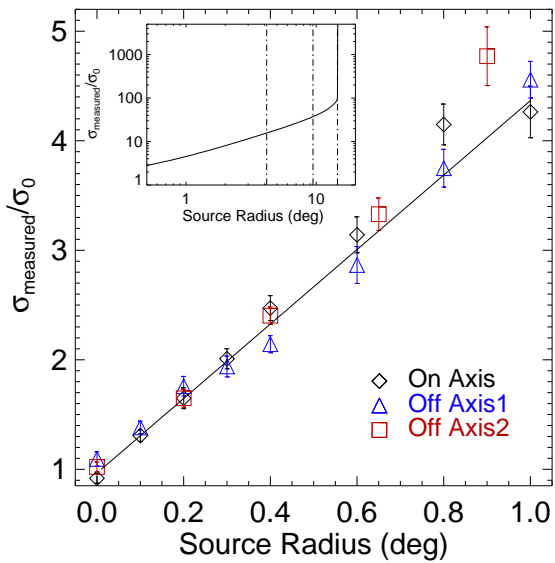}

\caption{The ratio between the measured error of our method and that
predicted for a point-like source ($\sigma_{0}$) as a function
of the source radius for the three relative source positions. The
solid line is given by the equation \ref{e:transfo_var1} and its
large-scale behavior up to the total \ibis FOV is shown in the
encapsulated plot.}

\label{f:simu_sou}
\end{figure}

Figure \ref{f:simu_sou} presents the ratio between the measured error
and that derived from the equation \ref{e:transfo_var1} for a point-like
source (labelled $\sigma_{0}$) as a function of the source radius for
the three distinct source positions. This ratio characterizes the loss
in sensitivity due to the source extent. The error bars were
calculated from the error of the fit performed on each reconstructed
fluxes histogram to estimate the error of the method. For any source
size and position, the dispersion is estimated to be less than 10 \%.


\section{Discussion}
\label{s:discussion}

We have shown that it is always possible to obtain an estimate of the flux of
any extended source with coded mask imaging. This is an intrinsic property of
these imaging systems where the source flux is obtained through a linear combination of
detector-pixel count rates. We have derived analytical expressions for estimating the
global flux and its associated error of any extended sky source which were tested by
Monte-Carlo simulations. This work was done relating to what one could
expect to observe in the \gammaray domain through a MURA coded mask instrument
such as \ibis onboard the \integ satellite. For \ibis, extended sources with
a radius greater than $\sim$ 4.5$^{\circ}$ will be confused with their associated ghosts.
One will need a specific method to clean the images but this is beyond the scope
of this paper. The advantage of this method, as long as the integration region contains
all the source, is that the observer does not have to assume any source geometry during the
analysis as is the case for methods using model fitting. Of course, in principle, the optimum
signal to noise ratio is attained with the smallest integration region still containing all
the source. However, due to the flux dilution, the geometry of the extended source is generally
not apparent in the maps and the observer may not always be provided with a precise source geometry.
Nevertheless, the signal to noise ratio does not decrease rapidly with the integration region size
once the source is fully contained in it. In order to check the global method on an astrophysical
case, we analyzed the \ibis/\isgri data on the \coma, a bright known cluster of galaxies
with an \xray size of order $30'$ diameter. We applied this method, increasing the integration
region until a maximum signal to noise ratio was reached. In that way, we were able to extract for
the first time the global intensity of this extended source as well as interesting constraints
on its morphology. The results are presented in a joint paper \cite{c:renaud06}. In a forthcoming
paper, this method will be also used on the \ibis/\isgri observations on the Vela Junior SNR
(Renaud \etal 2006, in preparation).

Usually, the observer needs a lot of integration time for detecting an extended source
($\sim$ 5 $\times$ 10$^{5}$ secondes for the \coma) and will have to construct mosaics
of individual images. After obtaining the individual images with the standard analysis,
the next step is to build mosaics as usual by weighting each image by their associated
variance and transform them following the practical tips on the use of the
method given in Appendix \ref{a:practical_tips}. Any observer interested in
this method may contact the first author to obtain the map $\tilde{A}$.


\begin{acknowledgements}

The authors thank the anonymous referee for his numerous and excellent
suggestions. The present work was partially based on observations with \integ
an ESA project with instruments and science data center (ISDC) funded by
ESA members states (especially the PI countries: Denmark, France,
Germany, Italy, Switzerland, Spain, Czech Republic and Poland, and
with the participation of Russia and the USA). \isgri has been
realized and is maintained in flight by CEA-Saclay/DAPNIA with the
support of CNES. E.~K. is supported by the European Comission through
a Marie Curie International Reintegration Grant (INDAM, Contract No
MIRG-CT-2005-017203). S.~P.~R. acknowledges support from NASA through
INTEGRAL observing award NAG5-13092.

\end{acknowledgements}

\newpage

\appendix

\section{Deconvolution Algorithm}
\label{s:appendixB}

In this section, we will adopt the following notations:

\begin{eqnarray}
(k,l)   & \xrightarrow[]{} &    {\rm detector} \\
(i,j)   & \xrightarrow[]{} &    {\rm real}\hspace{1mm}{\rm sky} \\
(m,n)   & \xrightarrow[]{} &    {\rm reconstructed}\hspace{1mm}{\rm sky}
\end{eqnarray}

D is the detector, S and S' the real and reconstructed skies, M
the coded mask, R the decoding matrix, and B the background. By
definition:

\begin{eqnarray}
D_{kl}              & = & \sum_{ij} S_{i,j}M_{i+k,j+l} + B_{k,l} \\
S'_{mn}             & = & \sum_{kl} R_{k+m,l+n}D_{kl}
\end{eqnarray}

\noindent with the following arrays:

\begin{eqnarray}
R_{k+m,l+n}    & = &   \frac{G^{+}_{k+m,l+n} - Bal_{mn}\times{G^{-}_{k+m,l+n}}}{E_{mn}} \\
Bal_{mn}       & = &   \frac{\sum_{kl} G^{+}_{k+m,l+n}}{\sum_{kl} G^{-}_{k+m,l+n}} \\
E_{mn}         & = &    \sum_{kl}G^{+}_{k+m,l+n} M_{k+m,l+n} - Bal_{mn}\sum_{kl} G^{-}_{k+m,l+n}M_{k+m,l+n}
\end{eqnarray}

\noindent where G$^{+}$, G$^{-}$ the decoding arrays related to the coded
mask M (see Goldwurm \etal 2003), Bal$_{m,n}$ is called the balance array
in order to have a flat reconstructed flux image and E$_{m,n}$ correponds
to the imaging efficiency to correct for partial exposure outside the FCFOV.

\subsection{Reconstructed fluxes}
\label{a:flux}

The standard reconstructed flux image is given by:

\begin{equation}\label{e:gen_flux}
S'_{mn} = \sum_{kl} R_{k+m,l+n} \sum_{ij} S_{ij}M_{i+k,j+l} + \sum_{kl} R_{k+m,l+n} B_{k,l}
\end{equation}

We propose to construct images of flux per steradian (per sky pixel)
I'$_{mn} = S'_{mn}/\tilde{A}_{mn}$ where $\tilde{A}_{mn}$ is the SPSF area.
The global flux of any source (point-like or extended) I'$_{S}$ is then
given by the sum of the intensities within a region $\Omega$ encompassing
the source extent. Therefore:

\begin{equation}\label{e:intensity}
I'_{S} = \sum_{\Omega} \frac{1}{\tilde{A}_{mn}} \sum_{ij} S_{ij}
\sum_{kl} R_{k+m,l+n} M_{i+k,j+l} + \sum_{\Omega}
\frac{1}{\tilde{A}_{mn}} \sum_{kl} R_{k+m,l+n} B_{k,l}
\end{equation}

\noindent Assuming that the detector D was well
background-substracted before the deconvolution, the second term
of the right-hand side of the above equation is null.

$\tilde{A}_{mn}$ represents the integrated normalized SPSF. By
definition, we have:

\begin{eqnarray}\label{e:spsf1}
SPSF_{ij}(m,n)  & = &   \sum_{kl} R_{k+m,l+n} M_{i+k,j+l} \\\label{e:spsf2}
\tilde{A}_{mn}  & = &   \sum_{m',n' \in K^{m,n}} SPSF_{mn}(m',n')
\end{eqnarray}

\noindent where SPSF$_{ij}$(m,n) is the SPSF value at the position
(m,n) of the source at the position (i,j) with a unit-flux.
K$^{m,n}$ defines the local region encompassing the SPSF of the
point source at the position (m,n). Substituting equation
\ref{e:spsf1} in \ref{e:intensity}:

\begin{eqnarray}
I'_{S} & = & \sum_{ij} S_{ij} \sum_{m,n \in \Omega} \frac{SPSF_{ij}(m,n)}{\tilde{A}_{mn}} \\
       & = & \sum_{ij} S_{ij} \sum_{m,n \in K^{i,j}} \frac{SPSF_{ij}(m,n)}{\tilde{A}_{mn}}
\end{eqnarray}

For each point source S$_{ij}$, we have reduced the sum over $\Omega$ (\ie the
source extent) to the sum over K$^{i,j}$, the local region around the point
source at (i,j). Thus, if we assume for each unresolved point that
$\tilde{A}_{mn} = \tilde{A}_{ij}$ is constant within K$^{i,j}$, we finally obtain:

\begin{equation}
I'_{S} \approx \sum_{ij} S_{ij}
\end{equation}

\subsection{Reconstructed variances}
\label{a:variance}

From equation \ref{e:gen_flux}, the reconstructed flux
variance is given by:

\begin{equation}\label{e:gen_var}
\sigma'^{2}_{S',mn} = \sum_{ij}\sigma^{2}_{S,ij}(\sum_{kl}R_{k+m,l+n}M_{i+k,j+l})^{2} + \sum_{kl}\sigma^{2}_{B,kl}R_{k+m,l+n}^{2}
\end{equation}

\noindent where $\sigma^{2}_{S,ij}$ is the variance related to the
celestial source at the position (i,j) and $\sigma^{2}_{B,kl}$
that related to the detector background. $\sigma'^{2}_{S',mn}$ is
equivalent to the OSA flux variance $\sigma'^{2}_{OSA}$.

In most cases in the high-energy domain, the variance due to the
celestial source(s) is very small compared to that due to the
background count rate. Therefore, in the calcul of the
variance associated to the transformation I'$_{S} = \sum_{\Omega}
S'_{mn}/\tilde{A}_{mn}$, the contribution of the first term in
equation \ref{e:intensity} can be neglected. Given that the
detector pixels are independent, we obtain:

\begin{equation}
\sigma'^{2}_{I'_{S}} = \sigma^{2}_{B}\sum_{kl} \left(\sum_{\Omega} \frac{R_{k+m,l+n}}{\tilde{A}_{mn}}\right)^{2}
\end{equation}

\noindent In the above equation, we have assumed for simplicity that the detector
variance $\sigma^{2}_{B,kl}$ is spatially constant. As the decoding
arrays G$^{+}$ and G$^{-}$ are given by (2M - 1)$^{+}$ and (2M - 1)$^{-}$,
respectively, we obtain:

\begin{eqnarray*}
\sigma'^{2}_{I'_{S}}    & \approx & 2\sigma^{2}_{B} \sum_{kl} \sum_{\Omega_1,\Omega_2}
       \frac{R_{k+m1,l+n1}(M^{+}_{k+m2,l+n2} - Bal_{m2,n2}M^{-}_{k+m2,l+n2})}
       {\tilde{A}_{m1,n1}\tilde{A}_{m2,n2}E_{m2,n2}}
\end{eqnarray*}
\begin{eqnarray}
            &    -    & \sigma^{2}_{B} \sum_{kl} \sum_{\Omega_1,\Omega_2}
       \frac{R_{k+m1,l+n1}(1^{+} - Bal_{m2,n2}\times{1^{-}})}{\tilde{A}_{m1,n1}
       \tilde{A}_{m2,n2}E_{m2,n2}}
\end{eqnarray}

\noindent In the FCFOV, the balance array is almost -1 and only varies within $\sim$ 5 \%
in the PCFOV. Therefore:

\begin{eqnarray}
\sigma'^{2}_{I'_{S}}    & \approx & 2\sigma^{2}_{B} \sum_{kl} \sum_{\Omega_1,\Omega_2}\frac{R_{k+m1,l+n1}.M_{k+m2,l+n2}}{\tilde{A}_{m1,n1}\tilde{A}_{m2,n2}E_{m2,n2}} \\
            &    -    & \sigma^{2}_{B} \sum_{kl} \sum_{\Omega_1,\Omega_2}\frac{R_{k+m1,l+n1}.1}{\tilde{A}_{m1,n1}\tilde{A}_{m2,n2}E_{m2,n2}}
\end{eqnarray}

The second term of the equation is assumed to be 0, since the deconvolution of any
flat image gives a reconstructed sky of 0 values. We have expanded the first term,
following the above definitions of $\tilde{A}$ and R, and finally found a simple
expression for the associated intensity variance such that:

\begin{eqnarray}\label{e:var_final1}
\sigma'^{2}_{I'_{S}}      & \approx &   \sigma^{2}_{B,ct.s^{-1}} \sum_{\Omega}\frac{1}{\tilde{E}_{mn}\tilde{A}_{mn}} \\
\sigma^{2}_{B,ct.s^{-1}}  &    =    &   \frac{\sum_{kl} D_{k,l}}{{\rm Time}^{2}} \\
\tilde{E}_{mn}        &    =    &   \frac{E_{mn}}{{\rm On-Axis \hspace{0.2mm} Aperture}}
\end{eqnarray}

\noindent In this expression, the final variance
$\sigma'^{2}_{I'_{S}}$ is in count s$^{-2}$ for homogeneity with
I'$_{S}$ (in count s$^{-1}$). The On-Axis Aperture is the number
of detector pixels which are illuminated through the mask by
an on-axis source ($\approx$ half of the \isgri camera, \ie 8192
pixels) and $\tilde{E}_{mn}$ is the imaging efficiency of the
telescope.

\section{Practical tips on the use of the method}
\label{a:practical_tips}

Equation \ref{e:var_final1} presents a simple expression of the
intensity variance associated with the method we developed for
reconstructing the flux of an extended source. In order to provide
the observer analyzing the \ibis data with an easy-to-use
expression, in relation to the standard OSA sky maps, we found that:

\begin{equation}\label{e:var_final2}
\sigma'^{2}_{OSA} \approx
\frac{\sigma^{2}_{B,ct.s^{-1}}}{\tilde{E}_{mn}}
\end{equation}

\noindent and then, from equation \ref{e:var_final1}:

\begin{equation}\label{e:var_final2bis}
\sigma'^{2}_{I'_{S}} = \sum_{\Omega}
\frac{\sigma'^{2}_{OSA}}{\tilde{A}_{mn}}
\end{equation}

\noindent If $\tilde{A}$ does not suffer from large variations
over $\Omega$, one may approximate the last equation:

\begin{equation}
\sigma'^{2}_{I'_{S}} \approx <\sigma'^{2}_{OSA}> \times{\frac{N_{pix}}{<\tilde{A}>}} = N_{spsf} <\sigma'^{2}_{OSA}>
\end{equation}

\noindent where $\sigma'^{2}_{OSA}$ is an estimate of the OSA flux variance averaged
over $\Omega$. The sum in equation \ref{e:var_final2bis} is
reduced to the number of sky pixels within $\Omega$ divided by $<\tilde{A}>$, \ie the number of
\ibis SPSFs. The transformation $S'_{OSA}\rightarrowtail{I'}$ allows the construction
of sky images where anybody may integrate intensities over any region $\Omega$ in order to
properly measure the global flux of an extended source. The associated variance is then given by
equation \ref{e:var_final2bis}. The following equations allow the user to construct
mosaics of a set of individual images:

\begin{eqnarray}\label{e:var_final3}
S'_{mosaic}     & = & \frac{\sum{S'_{OSA}/\sigma'^{2}_{OSA}}}{\sum{1/\sigma'^{2}_{OSA}}} \\
\tilde{A}_{mosaic}  & = & \frac{\sum{\tilde{A}/\sigma'^{2}_{OSA}}}{\sum{1/\sigma'^{2}_{OSA}}} \\
\sigma'^{2}_{mosaic}    & = & \frac{1}{\sum{1/\sigma'^{2}_{OSA}}}
\end{eqnarray}

\noindent As for an individual image, the global flux and the associated error of an
extended source will be given by:

\begin{eqnarray}\label{e:var_final4}
I'_{S}          & = & \sum_{\Omega} \frac{S'_{mosaic}}{\tilde{A}_{mosaic}} \\
\sigma'^{2}_{I'_{S}}    & = & \sum_{\Omega} \frac{\sigma'^{2}_{mosaic}}{\tilde{A}_{mosaic}}
\end{eqnarray}

\clearpage

\end{document}